\documentclass[11pt]{article}
\usepackage{latexsym}
\usepackage{amssymb}
\usepackage[dvips]{graphicx}
\usepackage{epsfig}
\usepackage{rotating}

\usepackage{amsfonts}
\usepackage{amsmath}
\input xy
\xyoption{all}

\newtheorem{definition}{Definition}[section]

\newtheorem{property}{Property}[section]

 \newtheorem{thm}{Theorem}[section]

 \newtheorem{prop}[thm]{Proposition}

 \numberwithin{equation}{section}
\newcommand{\beq}{\begin{equation}}
\newcommand{\eeq}{\end{equation}}
\newcommand{\bea}{\begin{eqnarray}}
\newcommand{\eea}{\end{eqnarray}}

\newcommand{\R}{\mathbb{R}}

\begin{document}
\begin{center}
{\LARGE \bf {$n-$ary  star  product:  construction and integral\\ 

\vspace{0.3cm}
representation}}

\vspace{1cm}
Mahouton Norbert Hounkonnou$^{1\dag}$ and Dine Ousmane
Samary$^{2,1*}$

\vspace{1cm}
 $^{1}${\em
International Chair in Mathematical Physics
and Applications}\\
{\em (ICMPA--UNESCO Chair), 072 B.P. 50  Cotonou, Republic of Benin}\\

\vspace{0.5cm}

$^2${\em Perimeter Institute for Theoretical Physics\\
31 Caroline St. N.
Waterloo, ON N2L 2Y5, Canada}

\vspace{0.5cm}

E-mails:   $^{\dag}$norbert.hounkonnou@cipma.uac.bj,\\
$^{*}$dsamary@perimeterinstitute.ca

\begin{abstract}
This paper addresses a construction of an  $n-$ary  star  product. Relevant  identities are given. Besides, the formalism is illustrated by a computation of eigenvalues and eigenfunctions for a physical system of  coupled oscillators in an $n$ dimensional phase space. 
\end{abstract}


\end{center}

\section{Introduction}
Several works were devoted to generalizations of Lie algebras to various types of $n$-ary algebras. To cite a few, see the works by Filippov, Hanlon, Vinogradov, Takhtajan and collaborators in  
 \cite{V. T. Filippov1}-\cite{P. Hanlon}-\cite{Takhtajan}.
In the same time, and
 intended to physical applications, the new  algebraic structures were considered in the case of the algebra $C^{\infty}(M )$ of functions on a $C^{\infty}-$manifold $M$, under the assumption that the operation is a derivation of each entry separately. In this way one got the Nambu-Poisson brackets, see  e.g. \cite{Ph. Gautheron}.
 The same versality was observed for generalized Poisson brackets  in  \cite{A. M. Perelomov } (and references therein)
 providing unexpected algebraic structures on vector fields, which played an essential role in the construction of universal enveloping algebras of Filippov algebras ($n$-Lie algebras). See, for instance, \cite{A. S. Dzhumadildaev} and references therein.
For many other applications, especially to theoretical physics, see a nice and interesting  survey
of $ n-$ary analogues of Lie algebras written by Azc\'arraga and Izquierdo \cite{Iz}.

This work intends to provide a construction of an $n-$ary star product, to  investigate some  identities related to it, and to give a concrete illustration on a physical system of coupled oscillators in an $n-$dimensional phase space.

The paper is organized as follows. In section \ref{sec1}, we focus on the study of a $3-$ary star product  in a $3$-dimensional Euclidean space.
We prove that the $3-$ary star product is  distributive, associative and satisfies the Jacobi identity. We also construct its integral representation  which should likely allow to establish new classes of 
solvable actions in the context  of field theories. 
Section \ref{sec2} is devoted to the generalization of this star product  in higher dimension, i. e. for $n>3$.  In section \ref{sec3} we provide a simple application of such a star product on a  physical system of  coupled oscillators for which the eignenvalues and eigenfunctions are explicitly  computed. 
In the  section \ref{sec4},  we give some  concluding remarks. 
\section{$3-$ary star product }\label{sec1}
\subsection{Definition and properties}
 We start by the  following definition:
\begin{definition}
 Consider $x=(x_1,x_2,x_3)\in\mathbb{R}^3$.
  Let   $\sigma_c$ be the cyclic permutation over the set
 $\{1,2,3\}$ such that  
$\sigma_c(1)=2,\,\sigma_c(2)=3,\,\sigma_c(3)=1,$ and 
 ${\mathcal A}= \Big ({\mathcal S}\Big(\R^3\Big), \star\Big)$  be the Schwartz space of  (smooth,
rapidly decreasing, together with all their derivatives, faster than the reciprocal of any polynomial at $\infty$) real valued
functions  on $\R^3$, endowed with a $3$-ary star product defined at a point $x$ as follows:
\begin{eqnarray}\label{dino}
\Big(f\stackrel{g}{\star}h\Big)(x):={\rm\bf m}\Big[e^{\mathcal{P}(\theta_1,\theta_2,\theta_3)}(f\otimes g\otimes h)(x)\Big]
\end{eqnarray}
where
\begin{equation}\label{lole}
\mathcal{P}(\theta_1,\theta_2,\theta_3)=\sum_{k=1}^3\frac{i\theta_k}{2}\Big(\partial_k\otimes \partial_{\sigma_c(k)}\otimes\partial_{
\sigma_c\circ\sigma_c(k)}
-\partial_k\otimes \partial_{\sigma_c\circ\sigma_c(k)}
\otimes\partial_{\sigma_c(k)}\Big),
\end{equation}
${\rm \bf m}(f\otimes g\otimes h)=fgh;$
  the parameters $\theta_i,\,i=1,2,3,$ are real numbers.  
\end{definition}
The conjugate of $ \Big(f\stackrel{g}{\star}h\Big)(x),$ denoted by 
$\overline{ \Big(f\stackrel{g}{\star}h\Big)(x)}$  is 

\beq
\overline{ \Big(f\stackrel{g}{\star}h\Big)(x)}:={\rm\bf m}\Big[e^{-\mathcal{P}(\theta_1,\theta_2,\theta_3)}(f\otimes g\otimes h)(x)\Big].
\eeq

Note that, if the space $\mathbb{R}^3$ is the space of  position coordinates, then  the parameters $\theta_i$ must be of order of magnitude of the Planck  volume, i.e. $[\theta_i]=[L^3]$. There exist  several  constructions of $n-$ary star products in the literature.  For instance in \cite{bghm}, see remark (1.1)  where a more  general star product with tensor $\Theta^{\mu_1\mu_2\cdots\mu_n}$ is examined. In \eqref{lole}, we adopt   a specific choice of $\Theta$ motivated by the fact that the resulting $n-$ary product has nice properties like the associativity, the distributivity,  a consistent Jacobi identity, and so on. Moreover, it well behaves in concrete applications like for the example of coupled oscillators exhibited in this work.

 Let $f_i,g_i,h_i \in {\mathcal A},\,\,i=1,2,3.$  The following properties are satisfied for the defined $3-$ary star product:
\begin{property}[Distributivity] The distributivity of the $3-$ary star product \eqref{dino} is 
given by the three following  relations:
\begin{eqnarray}\label{dino1}
 (f_1+f_2)\stackrel{g_1}{\star}h_1&=&f_1\stackrel{g_1}{\star}h_1+f_2\stackrel{g_1}{\star}h_1,\\
 f_1\stackrel{(g_1+g_2)}{\star}h_1&=&f_1\stackrel{g_1}{\star}h_1+f_1\stackrel{g_2}{\star}h_1,\\
f_1\stackrel{g_1}{\star}(h_1+h_2)&=&f_1\stackrel{g_1}{\star}h_1+f_1\stackrel{g_1}{\star}h_2.
\end{eqnarray}
\end{property}
\begin{property}[Associativity] The associativity of the $3-$ary star product \eqref{dino}
can be defined as: 
\begin{eqnarray}\label{mam}
(f_1\stackrel{g_1}{\star}h_1)\stackrel{g_2}{\star}h_2=f_1\stackrel{g_1}{\star}(h_1\stackrel{g_2}{\star}h_2).
\end{eqnarray}
\end{property}
Besides, it appears natural to define a $3-$ary star bracket as below:
\begin{definition}[$3-$ary star bracket]
\begin{eqnarray}
\{f,h\}_{\star,g}:=f\stackrel{g}{\star}h-h\stackrel{g}{\star}f
\end{eqnarray}
\end{definition}
with the following properties:
\begin{property}[Skew-symmetry]
\begin{eqnarray}
\{f,h\}_{\star,g}=-\{h,f\}_{\star,g}
\end{eqnarray}
\end{property}
\begin{property}[Jacobi identity] 
\bea
\{l,\{f,h\}_{\star,g_1}\}_{\star,g_2}+\{l,\{f,h\}_{\star,g_2}\}_{\star,g_1}+\{h,\{l,f\}_{\star,g_1}\}_{\star,g_2}\cr
\{h,\{l,f\}_{\star,g_2}\}_{\star,g_1}+\{f,\{h,l\}_{\star,g_1}\}_{\star,g_2}+\{f,\{h,l\}_{\star,g_2}\}_{\star,g_1}=0.
\eea
\end{property}
The proof of the Jacobi identity stems from the relation
\beq
\{l,\{f,h\}_{\star,g_1}\}_{\star,g_2}=(l\stackrel{g_2}{\star}f)\stackrel{g_1}{\star}h-(l\stackrel{g_2}{\star}h)\stackrel{g_1}{\star}f
-(f\stackrel{g_1}{\star}h)\stackrel{g_2}{\star}l+(h\stackrel{g_1}{\star}f)\stackrel{g_2}{\star}l.
\eeq
A thorough analysis of the defined star product and bracket properties is not concerned here and will be in the core of  a forthcoming work. Let us now sketch some interesting computations with this star product on the Euclidean coordinates. Indeed, the star composition of functions in ${\mathcal A}$ with coordinate functions gives rise to the following results:
\begin{prop}{\bf ($3-$ary star product of two functions in ${\mathcal A}$ and one coordinate function)}
\beq\label{an1}
x_k\stackrel{f}{\star}g=x_kfg+\frac{i\theta_k}{2}\Big(\partial_{\sigma_c(k)}f\partial_{\sigma_c
\circ\sigma_c(k)}g
-
\partial_{\sigma_c\circ\sigma_c(k)}f
\partial_{\sigma_c(k)}g\Big)
\eeq
\beq\label{an2}
g\stackrel{x_k}{\star}f=x_kfg-\frac{i\theta_{\sigma_c(k)}}{2}\partial_{\sigma_c(k)}g\partial_{\sigma_c\circ
\sigma_c(k)}f
+\frac{i\theta_{\sigma_c\circ
\sigma_c(k)}}{2}\partial_{\sigma_c\circ\sigma_c(k)}
g\partial_{\sigma_c(k)}f
\eeq
\beq\label{an3}
f\stackrel{g}{\star}x_k=x_kfg+\frac{i\theta_{\sigma_c(k)}}{2}\partial_{\sigma_c(k)}f\partial_{\sigma_c\circ
\sigma_c(k)}g
-\frac{i\theta_{\sigma_c\circ\sigma_c(k)}}{2}\partial_{\sigma_c\circ\sigma_c(k)}f\partial_{\sigma_c(k)}g.
\eeq
\end{prop}
\begin{prop}{\bf ($3-$ary star product of one function in ${\mathcal A}$ and two coordinate functions)}
\beq
x_k\stackrel{x_{\sigma_c(k)}}{\star}f=x_kx_{\sigma_c(k)}f+\frac{i\theta_k}{2}\partial_{\sigma_c\circ\sigma_c(k)}f
\eeq
\beq
 x_k\stackrel{f}{\star}x_{\sigma_c(k)}=x_kx_{\sigma_c(k)}f-
\frac{i\theta_k}{2}\partial_{\sigma_c\circ\sigma_c(k)}f
\eeq
\beq
x_k\stackrel{x_{\sigma_c\circ\sigma_c(k)}}{\star}f=x_kx_{\sigma_c\circ\sigma_c(k)}f
-\frac{i\theta_k}{2}\partial_{\sigma_c(k)}f
\eeq
\beq
 x_k\stackrel{f}{\star}x_{\sigma_c\circ\sigma_c(k)}=x_kx_{\sigma_c\circ\sigma_c(k)}f+\frac{i\theta_k}{2}\partial_{\sigma_c(k)}f.
\eeq
\end{prop}
Therefore the following interesting results hold for a $3-$ary star product of any function $f\in {\mathcal A}$ with two coordinate functions :
\begin {prop} [$3-$ary star product complex conjugation]
Provided \eqref{dino}, we have, $\forall f\in  {\mathcal A},$
\bea
\overline{x_k\stackrel{x_{\sigma_c(k)}}{\star}f}= x_k\stackrel{f}{\star}x_{\sigma_c(k)},\,\,\,\, \overline{x_k\stackrel{x_{\sigma_c\circ\sigma_c(k)}}{\star}f}=x_k\stackrel{f}{\star}x_{\sigma_c\circ\sigma_c(k)}.
\eea
\end{prop}
In opposite, when any  two functions $f, g \in  {\mathcal A}$ enter in the $3-$ary star product with a unique coordinate function, the star noncommutativity is clearly expressed, i.e.
\bea
x_k\stackrel{g}{\star}f\neq f\stackrel{g}{\star}x_k,\,\,\,\,x_k\stackrel{g}{\star}f\neq f\stackrel{g}{\star}x_k,\,\,\, g\stackrel{x_k}{\star}f\neq f\stackrel{x_k}{\star}g.\nonumber
\eea
Furthermore,  introducing 
 complex variables $a_{kl}$ and their conjugate $\bar{a}_{kl}$ by
\beq\label{bek}
a_{kl}=\frac{x_k+ix_l}{\sqrt{2}},\quad \bar a_{kl}=\frac{x_k-ix_l}{\sqrt{2}},\,\,\,\, k,l=1,2,3,\,\,l\neq k
\eeq 
 and using  the equations \eqref{an1} \eqref{an2} and \eqref{an3}, 
we  establish the relations given in the three next propositions:
\begin{prop}{\bf ($3-$ary star product of two functions in ${\mathcal A}$ and one complex coordinate function)}
\begin{eqnarray}
a_{ij}\stackrel{f}{\star}g&=&a_{ij}fg+\frac{i\theta_i}{4}\Big(\partial_{\sigma_c(i)}f\partial_{\sigma_c^2
(i)}g-
\partial_{\sigma_c^2(i)}f
\partial_{\sigma_c(i)}g\Big)\cr
&&-\frac{\theta_j}{4}\Big(\partial_{\sigma_c(j)}f\partial_{\sigma_c^2
(j)}g-
\partial_{\sigma_c^2(j)}f
\partial_{\sigma_c(j)}g\Big)\\
\bar a_{ij}\stackrel{f}{\star}g&=&\bar{a}_{ij}fg+\frac{i\theta_i}{4}\Big(\partial_{\sigma_c(i)}f\partial_{\sigma_c^2
(i)}g-
\partial_{\sigma_c^2(i)}f
\partial_{\sigma_c(i)}g\Big)\cr
&&+\frac{\theta_j}{4}\Big(\partial_{\sigma_c(j)}f\partial_{\sigma_c^2
(j)}g-
\partial_{\sigma_c^2(j)}f
\partial_{\sigma_c(j)}g\Big)
\end{eqnarray}
\end{prop}
\begin{prop}{\bf ($3-$ary star product of two functions in ${\mathcal A}$ and one complex coordinate function)}
\begin{eqnarray}
g\stackrel{f}{\star}a_{ij}&=&a_{ij}fg+
\frac{i\theta_{\sigma_c(i)}}{4}\partial_{\sigma_c(i)}g\partial_{\sigma_c^2(i)}f-
\frac{i\theta_{\sigma_c^2(i)}}{4}\partial_{\sigma_c^2(i)}g\partial_{\sigma_c(i)}
f\cr
&&-\frac{\theta_{\sigma_c(j)}}{4}\partial_{\sigma_c(j)}g\partial_{\sigma_c^2(j)}f+
\frac{\theta_{\sigma_c^2(j)}}{4}\partial_{\sigma_c^2(j)}g\partial_{\sigma_c(j)}f
\\
g\stackrel{f}{\star}\bar a_{ij}&=&\bar{a}_{ij}fg+\frac{i\theta_{\sigma_c(i)}}{4}\partial_{\sigma_c(i)}f\partial_{\sigma_c^2(i)}f-
\frac{i\theta_{\sigma_c^2(i)}}{4}\partial_{\sigma_c^2(i)}g\partial_{\sigma_c(i)}f\cr
&&+\frac{\theta_{\sigma_c(j)}}{4}\partial_{\sigma_c(j)}g\partial_{\sigma_c^2(j)}f-
\frac{\theta_{\sigma_c^2(j)}}{4}\partial_{\sigma_c^2(j)}g\partial_{\sigma_c(j)}f
\end{eqnarray}
\end{prop}
\begin{prop}{\bf ($3-$ary star product of two functions in ${\mathcal A}$ and one complex coordinate function)}
\begin{eqnarray}
f\stackrel{a_{ij}}{\star}g&=&a_{ij}fg-\frac{i\theta_{\sigma_c(i)}}{4}\partial_{\sigma_c(i)}f\partial_{\sigma_c^2(i)}g+
\frac{i\theta_{\sigma_c^2(i)}}{4}\partial_{\sigma_c^2(i)}f\partial_{\sigma_c(i)}g\cr
&&+\frac{\theta_{\sigma_c(j)}}{4}\partial_{\sigma_c(j)}f\partial_{\sigma_c^2(j)}g-
\frac{\theta_{\sigma_c^2(j)}}{4}\partial_{\sigma_c^2(j)}f\partial_{\sigma_c(j)}g\\
f\stackrel{\bar a_{ij}}{\star}g&=&\bar{a}_{ij}fg-\frac{i\theta_{\sigma_c(i)}}{4}\partial_{\sigma_c(i)}f\partial_{\sigma_c^2(i)}g+
\frac{i\theta_{\sigma_c^2(i)}}{4}\partial_{\sigma_c^2(i)}f\partial_{\sigma_c(i)}g\cr
&&-\frac{\theta_{\sigma_c(j)}}{4}\partial_{\sigma_c(j)}f\partial_{\sigma_c^2(j)}g+
\frac{\theta_{\sigma_c^2(j)}}{4}\partial_{\sigma_c^2(j)}f\partial_{\sigma_c(j)}g.
\end{eqnarray}
\end{prop}
Evidently,  any two of these last results  cannot be straightforwardly obtained by  complex conjugation of each other.  Indeed,  we well get:
 \bea
\overline{a_{ij}\stackrel{f}{\star}g}\neq \bar a_{ij}\stackrel{f}{\star}g,\quad 
 \overline{g\stackrel{f}{\star}a_{ij}}\neq  g\stackrel{f}{\star}\bar a_{ij},\quad a_{ij}\stackrel{f}{\star}g\neq g\stackrel{f}{\star}a_{ij},\quad \overline{f\stackrel{a_{ij}}{\star}g}\neq  f\stackrel{\bar a_{ij}}{\star}g.\nonumber
\eea

\subsection{Integral representation}
To construct  the integral representation  of the $3-$ary star product \eqref{dino},  consider $s,x\in\mathbb{R}^3$ and  the plane  wave function of the form:
\beq
\exp(isx)=\exp[i(s_1x_1+s_2x_2+s_3x_3)].
\eeq  
Then,
their  $3-$ary star  product gives
\beq
 e^{ikx}\stackrel{e^{iqx}}{\star}e^{irx}=e^{\sum_{j=1}^3\frac{\theta_j}{2}\big(k_jq_{\sigma_c(j)}r_{\sigma_c^2(j)}
-
k_jq_{\sigma_c^2(j)}r_{\sigma_c(j)}
\big)+i(k+q+r)x}.
\eeq
Defining the quantity $\Omega^{qr}_j$ by
\beq
\Omega^{qr}_j=q_{\sigma_c(j)}
r_{\sigma_c^2(j)}-
q_{\sigma_c^2(j)}r_{\sigma_c(j)},
\eeq
which satisfies the conditions
\beq
\Omega^{qr}_j=-\Omega^{rq}_j,\,\, p\Omega^{qr}=r\Omega^{pq}=q\Omega^{rp},
\eeq
 the integral representation of the $3-$ary star product  of functions can be expressed as follows:
\begin{eqnarray}
(f\stackrel{g}{\star}h)(x)&=&\int\,d^3k\,d^3q\,d^3r\,\tilde{f}(k)\tilde{g}(q)\tilde{h}(r)\,\big(e^{ikx}\stackrel{e^{iqx}}{\star}e^{irx}\big)\cr
&=&\frac{1}{(2\pi)^9}\int\,d^3k\,d^3q\,d^3r\,d^3y\,d^3z\,
d^3t
\,f(y)g(z)h(t)\cr
&\times&e^{\frac{1}{2}\sum_{j}\theta_jk_j\Omega^{qr}_j} e^{ik(x-y)}e^{iq(x-z)}e^{ir(x-t)}.
\end{eqnarray}
It can be simplified, using the identity
\begin{eqnarray}\label{intrep}
\int\,d^3k\,e^{ik(x-y-\frac{i\theta}{2}\Omega^{qr})}&=&(2\pi)^3\delta^{(3)}\Big(x-y-\frac{i\theta}{2}\Omega^{qr}\Big),
\end{eqnarray}
into the form
\begin{eqnarray}
(f\stackrel{g}{\star}h)(x)&=&\frac{1}{(2\pi)^6}\int\,d^3q\,d^3r\,d^3y\,d^3z\,d^3t\,f(y)g(z)h(t)\cr
&\times&\delta^{(3)}(x-y-\frac{i}{2}q\Omega^{r\theta}) e^{iq(x-z)}e^{ir(x-t)}.
\end{eqnarray}


\section{Generalization to $n-$ary star product}\label{sec2}
In this section, we deal with  the generalization of the $3-$ary star product \eqref{dino} into an $n-$ary star product.  
\begin{definition}
Consider $x=(x_1,x_2,\cdots, x_n)\in\mathbb{R}^n.$
 Let $\sigma_c$ be the cyclic permutation over the set $\{1,2,\cdots, d\},$ i.e.  $\sigma_c(k)=k+1,\, k+1\leq n$, and $\sigma_c(n)=1,$ and let ${\mathcal A}= \Big ({\mathcal S}\Big(\R^n\Big), \star\Big)$  be the Schwartz space of  (smooth,
rapidly decreasing, together with all their derivatives, faster than the reciprocal of any polynomial at $\infty$) real valued
functions  on $\R^n$, endowed with an $n$-ary star product defined at a point $x$ as follows:
\bea
\star\{\cdot,\cdot,\cdots,\cdot\}: &&\underbrace{\mathcal{A}\times\mathcal{A}\times\cdots \times\mathcal{A}}_{n \mbox{ times}}\longrightarrow \mathcal{A}\cr
&&(f_1,f_2,\cdots f_n)\longmapsto \star\{f_1,f_2,\cdots f_n\}:=\star\{ f_i\}_{i=1}^n
\eea
where
\begin{eqnarray}
&&\label{product1}\star\{ f_i\}_{i=1}^n(x)={\rm\bf m}\Big[e^{\mathcal{P}(\theta_1,\theta_2,\cdots,\theta_n)}(f_1\otimes f_2\otimes\cdots\otimes f_n)(x)\Big],\\
&&\label{Cglas}{\rm \bf m}(f_1\otimes f_2\otimes\cdots\otimes f_n)=\prod_{i=1}^nf_i,
\end{eqnarray}
and
\begin{eqnarray}
&&\mathcal{P}(\theta_1,\theta_2,\cdots,\theta_n)=\sum_{k=1}^n\frac{i\theta_k}{2}\Big(\partial_k\otimes \partial_{\sigma_c(k)} \otimes\cdots\otimes\partial_{\sigma_c^n(k)}\cr
&&-\partial_k\otimes \partial_{\sigma_c^{n-1}(k)}
\otimes\cdots\otimes \partial_{\sigma_c^{1}(k)}\Big).
\end{eqnarray}
\end{definition}
For $f_i,g_i\in{\mathcal A},\, i\in\mathbb{N},$ 
we obtain:
\begin{prop}[$n-$ary star product of  functions in ${\mathcal A}$ and  coordinate functions]
\begin{eqnarray}
&&\star\{ f_i,x_p,g_j\}_{i=1,\cdots m-1,j=m+1,\cdots n-1}=x_p\prod_{i=1}^{m-1}f_i
\prod_{j=m+1}^{n}g_j
\cr
&&+\frac{i\theta_{p-m+1}}{2}\prod_{i=1}^{m-1}\partial_{\sigma_c^{i-1}(i)} f_{i}\prod_{i=m+1}^{n}\partial_{\sigma_c^{i-1}(i)} g_{i}\cr
&&-\frac{i\theta_{p+m-n-1}}{2}\prod_{i=1}^{m-1}\partial_{\sigma_c^{n-i+1}(i)} f_{i}\prod_{i=m+1}^{n}\partial_{\sigma_c^{n-i+1}(i)} g_{i}
\end{eqnarray}
\end{prop}

\begin{prop}[$n-$ary star product of  functions in ${\mathcal A}$ and  coordinate functions]
\begin{eqnarray}
\star\{ f_i,x_p,g_j\}_{i=1,\cdots m-1,j=m+1,\cdots n-1}&=&x_p\prod_{i=1}^{m-1}f_i\prod_{j=m+1}^{n}g_j
+\frac{i\theta_{p-m+1}}{2}\prod_{i=1}^{m-1}\partial_{\sigma_c^{i-1}(i)} f_{i}\prod_{i=m+1}^{n}\partial_{\sigma_c^{i-1}(i)} g_{i}\cr
&-&\frac{i\theta_{p+m-n-1}}{2}\prod_{i=1}^{m-1}\partial_{\sigma_c^{n-i+1}(i)} f_{i}\prod_{i=m+1}^{n}\partial_{\sigma_c^{n-i+1}(i)} g_{i}.
\end{eqnarray}
\end{prop}

Therefore,  for $  p,q\in\{1,2,\cdots,n\},$ these relations show that any two of them cannot be given by complex conjugation. In fact,  we have:
\begin{eqnarray}
\overline{\star\{ a_{pq},f_i\}}_{i=1,\cdots,n-1}&\neq& \star\{ \bar a_{pq},f_i\}_{i=1,\cdots,n-1},\cr
 \overline{\star\{ f_i,a_{pq}\}_{i=1,\cdots,n-1}}&\neq& \star\{ f_i,\bar a_{pq}\}_{i=1,\cdots,n-1}\cr
\overline{\star\{ f_i,a_{pq},g_j\}}_{i=1,\cdots m-1,j=m+1,\cdots n-1}&\neq& \star\{ f_i,\bar a_{pq},g_j\}_{i=1,\cdots m-1,j=m+1,\cdots n-1}.\nonumber
\end{eqnarray}
The $n-$ary star product  integral representation,  for $n$ arbitrary points,  can be also computed 
by the same method as in the previous   section \ref{sec2}.  By
considering the plane wave functions 
\beq
e^{isx}=e^{i(s_1x_1+\cdots+s_nx_n)}, \quad s,x\in\mathbb{R}^n,
\eeq
the following result can be proved:

\begin{prop}[$n-$ary star product integral representation]
\begin{eqnarray}
\star\{f_i\}_{i=1}^n(x)&=&\frac{1}{(2\pi)^{2n}}\int\,\prod_{j=1}^{n-1}d^nq_j\prod_{j=1}^{n}d^ny_j
\,f(y_j)
\delta^{(n)}(x-y-\frac{i}{2}q\Omega^{r\theta}) \prod_{j=1}^{n-1}e^{iq(x-y_j)}.
\end{eqnarray}
\end{prop}

\section{Application }\label{sec3}
Consider   a physical system described by the Hamiltonian model:
\begin{eqnarray}\label{HH}
H&=&\sum_{j=1}^n x_j^2+\sum_{i<j=1}^n(\epsilon_{ij}
\lambda_{ij}x_ix_j)
+\sum_{i<j<k<l=1}^n(\epsilon_{ijkl}
\lambda_{ijk}x_ix_jx_kx_l)+\cdots\cr
&+&\sum_{i_1<i_2<\cdots<i_n=1}^n
(\epsilon_{i_1i_2\cdots
i_n}\lambda_{i_1\cdots i_n}x_{i_1}x_{i_2}\cdots x_{i_k})
\end{eqnarray} 
where $\lambda_{i_1\cdots i_k}, k\leq n,$ are the coupling constants and $\epsilon_{i_1i_2\cdots
i_k}$ the Levi-Civita  tensor of rank $k;$  $k=n$ if $n$ is even and $n-1$ if $n$ is odd. Using the orthogonal transformation
$\mathcal R$ such that $\mathcal R_{kl}x_l=X_k,$ 
the Hamiltonian 
 (\ref{HH})   can  be re-expressed in the new coordinates  $X$ as follows:
\begin{eqnarray}\label{HHH}
H=\sum_{i=1}^n\lambda^{(0)}_i  X_i^2+\sum_{i=1}^n\lambda^{(2)}_i  X_i^4+\cdots
\end{eqnarray}
allowing  a re-formulation  
  in terms of previous
 quantities  $a_{ij}$  and $\bar{a}_{ij},$ defined in \eqref{bek}, as follows:
\begin{eqnarray}
H=\sum_{i,j=1,i\neq j}^n\sum_{p=0}^{n-1}\Big(\lambda_i^{(2p)}
a_{ij}\bar{a}_{ij}\Big)^{p+1}.
\end{eqnarray}
For $\psi^{m}_k\in\mathcal{A}$, the eigenvalue problem  is given by the system of equations:
\begin{eqnarray}
 &&\star\{H,\psi^{m}_1,\cdots, \psi^m_{n-1}\}=E_{1,\bar n}\Big[\star\{1,\psi^{m}_1,\cdots, \psi^{m}_{n-1}\}\Big]\\
 &&\star\{\psi^{m}_1,H,\psi^{m}_2,\cdots, \psi^{m}_{n-1}\}=E_{2,\bar n}\Big[\star\{\psi^{m}_1,1,\psi^{m}_2,\cdots, \psi^{m}_{n-1}\}\Big]\\
\vdots\cr
 &&\star\{\psi^{m}_1,\psi^{m}_2,\cdots, \psi^{m}_{n-1},H\}=E_{n,\bar n}\Big[\star\{\psi^{m}_1,\psi^{m}_2,\cdots, \psi^{m}_{n-1},1\}\Big].
\end{eqnarray}
 $\bar n\in\mathbb{N}^n$ is a  n-vector  characterizing the quantum number associated to the Hamiltonian \eqref{HH} while 
 $\psi_j^m,\,\,j=1,2,\cdots n-1; \,m=1,2,\cdots n,$  are the eigenstates diagonalizing it.
The ground state satisfies  the equation
\begin{eqnarray}
\star\{a_{ij},\psi^0_1,\cdots,\psi^0_{n-1}\}=0
\end{eqnarray}
which can be explicitly solved  
to give:
\begin{eqnarray}
\psi_k^{0}=Ce^{-|x|^2/2}H_k(|x|^2/2)f(\lambda_k,|x|),\quad C\in\mathbb R
\end{eqnarray}
where $H_k$ is the Hermite polynomial;  the functions $f(\lambda_k,|x|):=f_k$ are orthogonal 
with the orthogonality  condition
\begin{eqnarray}\label{ortho}
\star \{f_{k_1},f_{k_2},\cdots f_{k_n}\}=C'\delta_{k_1k_2}\delta_{k_2k_3}
\cdots\delta_{k_{n-1}k_n}; \quad C'=C^n\in\mathbb{R}.
\end{eqnarray}
The case where $f_k=1,\,\forall k=1,2,\cdots n,$ is  solution of relation \eqref{ortho}. 

The excited states can be computed by   using the well-known harmonic oscillator algebraic  method performed with the raising operator, acting here as follows:
\begin{eqnarray}
\star\{\bar a_{ij},\psi^0_1,\cdots,\psi^0_{n-1}\}=f_1(n)\big(\star\{1,\psi^1_1,\cdots,\psi^1_{n-1}\}\big),
\end{eqnarray} 
where $f_1(n)$ is a function depending on the parameter $n$.

There result the following  expressions for the  eigenvalues and eigenfunctions characterizing  the considered physical model: 
\begin{eqnarray}
E_{k,\bar n}=\theta_k\Big(\lambda_k^{(0)}|\bar n|+\sum_{i=1}^n\lambda^{(k)}_i |\bar n|^2+\sum_{i,j=1}^n
\lambda^{(k)}_i\lambda^{(k)}_j|\bar n|^3+\cdots+\frac{n}{2}\Big)
\end{eqnarray}
and
\begin{eqnarray}
\psi_k^{ n}=Ce^{-|x|^2/2}H^{| n|}_k(|x|^2/2)f(\lambda_k,|x|).
\end{eqnarray}
$H^{|n|}_k$ stands for the  $(n,k)$-order Hermite polynomial.

\section{Concluding remarks}\label{sec4}
In this work, we have given a method of  constructing an  $n-$ary  star  product. Relevant  identities have been provided and discussed. A physical problem  of coupled oscillators has been treated. The associated eigenvalues and eigenfunctions have been explicitly computed.

\subsection*{Acknowledgment}
This work is partially supported by the ICTP through the OEA-ICMPA-Prj-15. The ICMPA is in partnership with the Daniel Iagolnitzer Foundation
(DIF), France.  Discussions with Joseph Ben Geloun are gratefully acknowledged. 
D.O  Samary was
supported in part by Perimeter Institute for Theoretical Physics and Fields Institute  for Research in Mathematical Sciences (Toronto). Research at Perimeter Institute is supported by the
Government of Canada through Industry Canada and by the Province of Ontario through the Ministry of Research
and Innovation.

\end{document}